\documentclass[onecolumn, usenatbib,useAMS]{mn2e}

\usepackage{natbib}
\usepackage{amsbsy}
\usepackage{amssymb}
\usepackage{amsmath}
\usepackage{graphicx}
\usepackage[justification=centering]{caption}
\usepackage{mathrsfs}
\usepackage[caption = false]{subfig}
\usepackage{array}

\usepackage [english]{babel}
\usepackage [autostyle, english = american]{csquotes}
\usepackage{natbib}
\MakeOuterQuote{"}

\date{}
\def\be{\begin{equation}}
\def\ee{\end{equation}}

\def\S{\mathcal {S}}
\def\A{\mathcal {A}}

\def\k{\mathbf{k}}
  
\def\kpar{k_{\parallel}}

\begin{document}

\title[BAO probe of Quintessence with lensing-21 cross-correlation]{ Probing Quintessence using BAO imprint on the cross-correlation of  weak lensing and post-reionization HI 21 cm signal}

%\author[Dash, Sarkar]{ Chandrachud B.V. Dash$^1$ \& Tapomoy Guha Sarkar$^{1}$ \\
%  \\
%  $^1$Department of Physics, Birla Institute of Technology and Science, Pilani Campus, %Pilani, 
%Jhunjhunu 333031, Rajasthan, India}
\author[Dash, Sarkar]{
Chandrachud B.V. Dash,$^{1}$\thanks{E-mail: cb.vaswar@gmail.com}
Tapomoy Guha Sarkar,$^{1}$\thanks{E-mail: tapomoy1@gmail.com}
\\
% List of institutions
$^{1}$Birla Institute of Technology \& Science, Pilani, Rajasthan India\\
}

% These dates will be filled out by the publisher
%\date{Accepted XXX. Received YYY; in original form ZZZ}

% Enter the current year, for the copyright statements etc.
\pubyear{2021}

\maketitle

\begin{abstract}
In this work we investigate the possibility of constraining a thawing
Quintessence scalar field model for dark energy.  We propose using the
imprint of baryon acoustic oscillation (BAO) on the cross-correlation
of post-reionization 21-cm signal and galaxy weak lensing convergence
field to tomographically measure the angular diameter distance
$D_A(z)$ and the Hubble parameter $H(z)$.  The projected errors in
these quantities are then used to constrain the Quintessence model
parameters.  We find that independent $600$hrs radio interferometric
observation at four observing frequencies $916 $MHz, $650$ MHz, $520$
MHz and $430 $MHz with a SKA-1-Mid like radio telescope in
cross-correlation with a deep weak lensing survey covering half the
sky may measure the binned $D_A$ and $H$ at a few percent level of
sensitivity. The Monte Carlo analysis for a power law thawing Quientessence
 model gives the $1-\sigma$ marginalized  bounds on the initial slope $\lambda_i$, 
  dark energy density parameter $\Omega_{\phi 0}$ and the shape of the potential 
  $\Gamma$ at $8.63\%$, $10.08\%$ and $9.75\%$ respectively.
The constraints improve to $7.66\%$, $4.39\%$ and $5.86\%$ 
   respectively when a joint analysis with SN and other probes is performed.

 \end{abstract}

\section{Introduction}
 Several decades of independent observations \citep{Perlmutter_1997,
   riess1998observational,bamba2012dark} confirm that our Universe is
 currently in an accelerated expansion phase. The cause of such cosmic
 acceleration is attributed to the so called "Dark energy",
 \citep{sahni2000case,RevModPhys.75.559,copeland2006dynamics,amendola_tsujikawa_2010}
 a fluid that violates the strong energy condition.  Einstein's
 cosmological constant ($\Lambda$) with an effective fluid equation of
 state (EoS) $ P/\rho = w(z) = -1$ provides the simplest explanation
 for the cosmic acceleration. While, several cosmological observations
 are consistent with the concordance LCDM model, there are several
 inconsistencies from both theoretical considerations (like smallness
 of $\Lambda$, the `fine tuning problem'), and observations (like the
 low redshift measurements of $H_0$ \citep{Riess_2016}). This has led
 to many significant efforts in developing alternate scenarios to
 model dark energy and thereby explaining the cosmic acceleration
 without requiring a cosmological constant.

Generally speaking there are two ways to tackle the problem. One
approach involves modifying the gravity theory itself on large scales
\citep{amendola_tsujikawa_2010}.  $f(R)$ modification to the Einstein
action
\citep{Khoury_2004,Starobinsky_2007,Hu_2007,nojiri2007introduction}
belongs to this approach of modeling cosmic acceleration. In a second
approach the matter sector of Einstein's field equation is modified by
considering a dark energy fluid with some nontrivial dynamics. In both
the approaches one may find an effective dark energy EoS which
dynamically varies as a function of redshift and in principle can be
distinguished from the cosmological constant ($\Lambda$). There are
many models for dark energy that predict a dynamical equation of
state. For example, in the quintessence models, dark energy arises
from a time dependent scalar field, $\phi$
\citep{Ratra-Peebles_1988,Steinhardt_1998,PhysRevD.59.123504,PhysRevLett.82.896,scherrer2008thawing}. However
these models still require fine tuning for consistency with
observations. A wide variety of phenomenological potentials have been
explored for quintessence field to achieve $w \approx -1$. In all
these models, the minimally coupled scalar field is expected to slowly
roll in the present epoch. However, other than a few restricted class
of potentials, it is difficult to prevent corrections from various
symmetry breaking mechanisms which tends to spoil the slow roll
condition \citep{panda2011axions}.

Weak gravitational lensing by intervening large scale structure
distorts the images of distant background galaxies. This is attributed
to the deflection of light by the fluctuating gravitational field
created by the intervening mass distribution and is quantified using
shear and convergence of photon geodesics.  The statistical properties
of these distortion fields are quantified using the shear/convergence
power spectrum. These imprint the power spectrum of the intervening
matter field, as well as cosmological evolution and thereby carries
the signatures of structure formation.  Dark energy affects the growth
of cosmic structures and geometric distances, which crucially affects
the power spectrum of the lensing distortion fields.  Thus, weak
lensing has become one of the important cosmological probes. Several
weal lensing experiments are either on-going or are upcoming, such as
the Dark Energy Survey \citep{abbott2016dark}, the Hyper Suprime-Cam
survey \citep{aihara2018hyper}, the Large Synoptic Survey Telescope
\citep{ivezic2008large}, the WideField Infrared Survey Telescope
\citep{wright2010wide,spergel2015wide}, and the Euclid
\citep{laureijs2011euclid}.

The 3D tomographic imaging of the neutral hydrogen (HI) distribution
is one of the promising tool to understand large scale structure
formation and nature of dark energy \citep{poreion1, poreion0}. The
dominant part of the low density hydrogen gets completely ionized by
the end of reionization around z $\sim$ 6
\citep{Gallerani_2006}. However, a small fraction of HI survives the
complex processes of reionization and is believed to remain housed in
the over-dense regions of IGM. These clumpy HI clouds remain neutral
amidst the radiation field of background ionizing sources as they are
self shielded and are the dominant source of the 21-cm radiation in
post-reionization epoch. Intensity mapping of such redshifted 21-cm
radiation aims to map out the large scale HI distribution without
resolving the individual DLA sources and promises to be a powerful
probe of large scale structure and background cosmological evolution
\citep{param1, param2, param3, param4}.  Several radio telescopes like
the GMRT \footnote{http://gmrt.ncra.tifr.res.in/}
OWFA\footnote{https://arxiv.org/abs/1703.00621},
MEERKAT\footnote{http://www.ska.ac.za/meerkat/},
MWA\footnote{https://www.mwatelescope.org/},
CHIME\footnote{http://chime.phas.ubc.ca/}, and
SKA\footnote{https://www.skatelescope.org/} are in the pursuit of
detecting the cosmological 21-cm signal for a tomographic imaging
\citep{Mao_2008}.

We consider the cross-correlation of HI 21-cm signal with the galaxy
weak lensing convergence field. It is known that
\citep{fonseca2017probing} cross-correlations of
individual tracers of IGM often offer crucial advantages over
auto-correlations. The systematic noise that arises in the individual
surveys is pose less threat in the cross-correlation signal as they
appear in the variance. Further, the foregrounds and contaminants of
individual surveys are, in most cases, uncorrelated and hence do not
bias the cross-correlation signal
\citep{GSarkar_2010,Vallinotto_2009}.  The cross-correlation of the
post-reionization HI 21 cm signal has been extensively studied
\citep{Sarkar_2009,Guha_Sarkar_2010,GSarkar_2010,Sarkar_2019,Dash_2021}.

The acoustic waves in the primordial baryon-photon plasma are frozen
once recombination takes place at $z \sim 1000$. The sound horizon at
the epoch of recombination provides a standard ruler which can be then
used to calibrate cosmological distances.  Baryons imprint the
cosmological power spectrum through a distinctive oscillatory
signature \citep{White_2005, Hu-eisen}.  The BAO imprint on the 21-cm
signal has been studied
\citep{sarkar2013predictions,sarkar2011imprint}.  The baryon acoustic
oscillation (BAO) is an important probe of cosmology
\citep{Eisenstein_2005, Percival_2007, Anderson_2012,
  shoji2009extracting, sarkar2013predictions} as it allows us to
measure the angular diameter distance $D_A(z)$ and the Hubble
parameter $H(z)$ using the the transverse and the longitudinal
oscillatory features respectively thereby allowing us to put stringent
constraints on dark energy models. We propose the BAO imprint on the
cross-correlation of 21-cm signal and weak lensing convergence as a
probe of Quintessence dark energy.

The paper is organized as follows. In Section-2 we discuss the
cross-correlation of weak lensing shear/convergence and HI excess
brightness temperature. We also discuss the BAO imprint and estimation
of errors on the BAO parameters namely the expansion rate $H(z)$,
angular diameter distance $D_A(z)$ and the dilation factor $D_V(z)$
from the tomographic measurement of cross-correlation power spectrum
using Fisher formalism. In Section-3 we discuss the background and
structure formation in quintessence dark energy models and constrain
the model parameters using Markov Chain Monte Carlo (MCMC)
simulation. We discuss our results and other pertinent observational
issues in the concluding section.
 
 \section{The cross-correlation signal}
 
Weak gravitational lensing \citep{Bartelmann_2001} by intervening
large scale structure distorts the images of distant background
galaxies. This is caused by the deflection of light by the fluctuating
gravitational field created by the intervening mass distribution
\citep{takada2004cosmological}. Weak lensing is a powerful
cosmological probe as galaxy shear is sensitive to both spacetime
geometry and growth of structures. The Weak-lensing convergence field
on the sky is given by a weighted line of sight integral
\citep{waerbeke2003} of the matter overdensity field $\delta$ as \be
\kappa({\vec \theta}) = \int_0^{\chi_s} ~ \A_{\kappa} (\chi)
\delta(\chi \vec \theta, \chi) d\chi \ee where $\chi_s$ is the maximum
distance to which the sources are distributed and the
cosmology-dependent function $\A_{\kappa}(\chi)$ is given by \be
\A_{\kappa} (\chi) = \frac{3}{2} \Omega_{m0}H_0^{2}
\frac{\chi}{{a(\chi)}} \int_0^{\chi_{_{s}}} n_{s}(z) \frac{dz}{d\chi
  '} \frac{\chi ' - \chi}{\chi '} d\chi ' \ee where $\chi$ denotes the
comoving distance and $a(\chi)$, the cosmological scale factor.  The
redshift selection function of source galaxies, $n_s(z)$ tends to zero
at both low and high redshifts. It is typically modeled as a peaked
function \citep{takada2004cosmological}, parametrized by ($\alpha$,
$\beta$, $z_0$) of the from \be n_s(z) = {N_0} z^{\alpha} e^{- \left(
  \frac{z}{z_0} \right)^{\beta}} \ee and satisfies the normalization
condition \be \int_0^\infty dz ~n_s(z) dz = \bar{n}_g \ee where
$\bar{n}_g$ is the the average number density of galaxies per unit
steradian.

 On large scales the redshifted HI 21-cm signal from post reionization
 epoch ($z<6$) known to be biased tracers of the underlying dark
 matter distribution \cite{Bagla_2010, Guha_Sarkar_2012, Sarkar_2016}.
 We use $\delta_T$ to denote the redshifted 21-cm brightness
 temperature fluctuations.  The post reionization HI signal has been
 studied extensively \citep{poreion0, poreion1, poreion2, poreion3,
   poreion4, poreion6, poreion7, poreion12, poreion8}.  We follow the
 general formalism for the cross-correlation of the 21-cm signal with
 other cosmological fields given in (\cite{Dash_2021}).  Usually for
 the investigations involving the 21-cm signal the the radial
 information is retained for tomographic study. The weak-lensing
 signal, on the contrary consists of a line of sight integral whereby
 the redshift information is lost. We consider an average over the
 21-cm signals from redshift slices and thus lose the individual
 redshift information but improve the signal to noise ratio when
 cross-correlating with the weak-lensing field.

We define a brightness temperature field on the sky by integrating
$\delta_T (\chi \bf{\hat{n}},\chi)$ along the radial direction as \be
\label{eq:summing-of-signal}
T(\hat{n}) = {\frac{1}{\chi_2 -\chi_1} \sum_{\chi_1}^{\chi_2 }\delta_T (\chi \bf{\hat{n}},\chi)} \Delta \chi
\ee
where $\chi_1$ and $\chi_2$ are the comoving distances corresponding to the redshift slices of the 21-cm observation over which the signal is averaged. 

Radio interferometric observations of the redshifted 21-cm signal
directly measures the complex Visibilities which are the Fourier
components of the intensity distribution on the sky. The radio
telescope typically has a finite beam which allows us to use the \lq
flat-sky' approximation. Ideally the fields $\kappa$ and $\delta_T$
are expanded in the basis of spherical harmonics. For convenience, we
use a simplified expression for the angular power spectrum by
considering the flat sky approximation whereby we can use the Fourier
basis. Using this simplifying assumption, we may approximately write
the cross-correlation angular power spectrum as \citep{Dash_2021} \be
\label{eq:crosssignal}
\nonumber
C^{ T \kappa}_\ell = \frac{1 }{\pi(\chi_2- \chi_1)} \sum_{\chi_1}^{\chi_2} \frac{\Delta \chi}{\chi^2} ~ \A_T (\chi) \A_\kappa (\chi) D_{+}^2 (\chi)  \int_0^{\infty} d\kpar \left [ 1 + \beta_T(\chi) \frac{\kpar^2}{k^2} \right ] P (k) 
\ee
where
  $k = \sqrt{\kpar^2 + \left ( \frac{\ell}{\chi} \right )^2 } $, $ D_{+}$ is the growing mode of density fluctuations, 
and $\beta_{T} = f/b_T$ is the redshift distortion factor - the ratio of the logarithmic growth rate  $f$ and the bias function and $b_T(k,z)$. The redshift dependent function $\A_{T}$ is given by \citep{bali,datta2007multifrequency,Guha_Sarkar_2012}
\be
\A_{T} = 4.0 \, {\rm {mK}} \,
b_{T} \, {\bar{x}_{\rm HI}}(1 + z)^2\left ( \frac{\Omega_{b0}
  h^2}{0.02} \right ) \left ( \frac{0.7}{h} \right) \left (
\frac{H_0}{H(z)} \right) 
\label{eq:21cmkernel}
\ee The quantity $b_T(k, z)$ is the bias function defined as ratio of
HI-21cm power spectrum to dark matter power spectrum $b_T^2 =
P_{HI}(z)/P(z)$.  In the post-reionization epoch $z<6$, the neutral
hydrogen fraction remains with a value $ {\bar{x}_{\rm HI}} = 2.45
\times 10^{-2}$ (adopted from \citet{ Noterdaeme_2009,Zafar_2013}).
The clustering of the post-reionization HI is quantified using $b_T$.
On sub-Jean's length, the bias is scale dependent
\citep{fang}. However, on large scales the bias is known to be
scale-independent. The scales above which the bias is linear, is
however sensitive to the redshift.  Post-reionization HI bias is
studied extensively using N-body simulations \citep{Bagla_2010,
  Guha_Sarkar_2012, Sarkar_2016, Carucci_2017}.  These simulations
demonstrate that the large scale linear bias increases with redshift
for $1< z< 4$ \citep{Mar_n_2010}. We have adopted the fitting formula
for the bias $b_T(k, z)$ as a function of both redshift $z$ and scale
$k$ \citep{Guha_Sarkar_2012, Sarkar_2016} of the post-reionization
signal as \be
\label{eqn:bias}
b_{T}(k,z) = \sum_{m=0}^{4} \sum_{n=0}^{2} c(m,n) k^{m}z^{n}
\ee
The coefficients  $c(m,n)$ in the fit function are adopted from  \citet{Sarkar_2016}.

 The angular power spectrum for two redshifts is known to decorrelate very fast in the radial direction \citep{poreion7}. We consider the summation in Eq (\ref{eq:summing-of-signal}) to extend over redshift slices whose separation is more than the typical decorrelation length. This ensures that in the computation of noise for each term in the summation may be thought of as an independent measurement and the mutual covariances between the slices may be ignored.

\subsection{The Baryon acoustic oscillation in the angular power spectrum}
The sound horizon at the epoch of recombination is given by \be
s(z_{d}) = \int_{0}^{a_r} \frac{c_s da}{a^2 H(a)} \ee where $a_r$ is
the scale factor at the epoch of recombination (redshift $z_d$) and
$c_s$ is the sound speed given by $c_s(a) = c/
\sqrt{3(1+3\rho_b/4\rho_\gamma)}$ where $\rho_b$ and $\rho_\gamma$
denotes the baryonic and photon densities respectively. The WMAP
5-year data constrains the value of $z_{d}$ and $s(z_d)$ to be $z_{d}
= 1020.5 \pm 1.6$ and $s (z_{d}) = 153.3 \pm 2.0 $Mpc
\citep{Komatsu_2009}. We shall use these as the fiducial values in our
subsequent analysis.  The standard ruler \lq$s$' defines a transverse
angular scale and a redshift interval in the radial direction as \be
\theta_s (z) =\frac{s(z_d)} {(1+z) D_A (z)} ~~~~~~~ \delta z_s =
\frac{s (z_d) H(z)}{c} \ee Measurement of $\theta_s$ and $\delta z_s$,
allows the independent determination of $D_A (z)$ and $H(z)$. The BAO
feature comes from the baryonic part of $P(k)$. Hence we isolate the
BAO power spectrum from cold dark matter power spectrum through $P_b
(k) = P(k) - P_c(k)$. The baryonic power spectrum can be written as
\citep{hu1996small, seo2007improved} \be
\label{eq:baops}
P_b (k) = A \frac{\sin x}{x} e^{-(k\sum_s)^{1.4}}e^{-k^2
  \sum_{nl}^2/2} \ee where $A$ is a normalization, $\sum_s =
1/k_{silk}$ and $\sum_s = 1/k_{nl}$ denotes the inverse scale of \lq
Silk-damping' and \lq non-linearity' respectively. In our analysis we
have used $k_{nl} = (3.07 h^{-1}Mpc)^{-1} $and $k_{silk} = (8.38
h^{-1}Mpc)^{-1} $ from \citet{seo2007improved} and $x =
\sqrt{k_\perp^2 s_\perp^2 + k_\parallel^2 s_\parallel^2}$.  We also
use the combined effective distance $D_V(z)$ defined as
\citep{Eisenstein_2005} \be D_V(z) \equiv \left[ (1+z)^2 D_A^2(z)
  \frac{c z}{H(z)} \right]^{1/3} \ee The changes in $D_A$ and $H(z)$
are reflected as changes in the values of $s_\perp$ and $s_\parallel$
respectively, and the errors in $s_\perp$ and $s_\parallel$
corresponds to fractional errors in $D_A$ and $H(z)$ respectively. We
use $p_1 = \ln (s^{-1}_{\perp})$ and $p_2 = \ln (s_{\parallel})$ as
parameters in our analysis.  The Fisher matrix is given by \be F_{ij}
= \sum_\ell \frac{1}{\sigma_{_{T \kappa}}^2} \frac{1 }{\pi(\chi_2-
  \chi_1)} \sum_{\chi_1}^{\chi_2} \frac{\Delta \chi}{\chi^2} ~ \A_T
(\chi) \A_\chi (\chi) D_{+}^2 (\chi) \int_0^{\infty} d\kpar \left [ 1
  + \beta_T(\chi) \frac{\kpar^2}{k^2} \right ] \frac{\partial P_b
  (k)}{\partial p_i} \frac{\partial P_b (k)}{\partial p_j} \ee \be =
\sum_\ell \frac{1}{\sigma_{_{T \kappa}}^2} \frac{ \A_T (\chi) \A_\chi
  (\chi)}{\pi(\chi_2- \chi_1)} \frac{\Delta \chi}{\chi^2} ~ D_{+}^2
(\chi) \int_0^{\infty} d\kpar \left [ 1 + \beta_T\frac{\kpar^2}{k^2}
  \right ] \left( \cos x - \frac{\sin x}{x} \right) f_i f_j ~A
e^{-(k\sum_s)^{1.4}}e^{-k^2 \sum_{nl}^2/2} \ee where $f_1 =
k_\parallel^2 / k^2 -1$, $f_2 = k_\parallel^2 / k^2 $ and $k^2 =
k_\parallel^2 + \ell^2 /\chi^2$. The variance $\sigma_{_{T \kappa}}$
is given by \be
\label{eq:variance}
\sigma_{_{T \kappa}}= \sqrt{ \frac {{(C^{\kappa}_\ell + N^{\kappa}_\ell )(
    C^{T}_\ell +  N^T_\ell )}}{   ({2\ell + 1}) f_{sky} }}  
\ee
 where $C^{\kappa}_\ell$ and $C^{T}_\ell $ are the convergence and 21-cm  auto-correlation angular power spectra respectively and $N^{\kappa}_\ell $ and  $N^T_\ell$ are the corresponding noise power spectra.
 
 The auto-correlation power spectra are given by (\cite{Dash_2021}) 
  \be
\label{eq:autoT}
C^{ T}_\ell = \frac{1 }{\pi(\chi_2- \chi_1)^2} \sum_{\chi_1}^{\chi_2} \frac{\Delta \chi}{\chi^2} ~ \A_T (\chi) ^2 D_{+}^2 (\chi)  \int_0^{\infty} d\kpar \left [ 1 + \beta_T(\chi) \frac{\kpar^2}{k^2} \right ]^2 P (k) 
\ee
\be
\label{eq:autok}
C^{ \kappa}_\ell = \frac{1 }{\pi}  \int_0^{\chi_s} \frac{d\chi}{\chi^2} ~ \A_\kappa (\chi)^2 D_{+}^2 (\chi)  \int_0^{\infty} d\kpar P (k) 
\ee
The noise is the convergence power spectrum is dominated by Poisson noise. Thus 
$N^\kappa = \sigma^2_\epsilon / \bar{n_g}$ where $\sigma_{\epsilon}$ is the galaxy-intrinsic rms
shear \citep{Hu_1999}. 
The source galaxy distribution is modeled using $( \alpha, \beta, z_0 ) = (1.28,~ 0.97,~ 0.41 )$ which we have adopted from \citet{chang2013effective}.
For the survey under consideration, we have taken $\sigma_{\epsilon}  = 0.4$ \citep{takada2004cosmological}.
 We use a visibility correlation approach to estimate the noise power spectrum $N^T_\ell$ for the 21-cm signal \citep{geil2011polarized,villaescusa2014modeling,Sarkar_2015}. 
 \be 
N^{T}_\ell = \left(\frac{ T^2_{sys} \lambda^2}{A_e}\right)^2 \frac{B}{T_{o}N_b(U,\nu)}
\ee
where $T_{sys}$ is the system temperature, $B$ is the total frequency bandwidth, $U = \ell / 2 \pi$,  $T_{o}$ is the total observation time, and $\lambda$ is the observed wavelength corresponding to the observed frequency $\nu$ of the 21 cm signal. The quantity $A_e$ is the effective collecting area of an individual antenna which can be written $A_e=\epsilon  \pi (D_{d}/2)^2$, where $\epsilon$ is the antenna efficiency and $D_{d}$ is the diameter of the dish. The $N_b (U,\nu)$ is the number density of baseline $U$ and can be expressed as
\be 
N_b (U,\nu) = \frac{N_{ant}(N_{ant}-1)}{2}\rho_{_{2D}}(U,\nu) \Delta U
\ee
where $N_{ant}$ is the total number of antenna in the radio array and $\rho_{_{2D}}(U,\nu)$ is the normalized baseline distribution function which follows the normalization condition $\int d^2U \rho_{_{2D}} (U,\nu)=1$. 
 The system temperature $T_{sys}$ can be written as a sum of contributions from sky and the instrument as
 \be T_{sys} = T_{inst} + T_{sky} \ee
 where
 \be 
 T_{sky} = 60K \left ( \frac{\nu}{300 \rm MHz} \right)^{-2.5} \ee
We consider a radio telescope with an operational frequency range of $400-950$ MHz.  We consider $200$ 
dish antennae in a radio interferometer roughly mimicking SKA1-Mid. The telescope parameters are summarized in table (\ref{tab:noise-params}).  The full frequency range is divided into $4$ bins centered on $916$ MHz, $650$ MHz, $520$ MHz and $430$MHz and $32$ MHz bandwidth each. To calculate the normalized baseline distribution function we have assumed that baselines are distributed such that the antenna distribution falls off as $ 1/r^2$.
We also assume that there is no baseline coverage below $30$m. We have also assumed $\Delta U  = A_e / \lambda^2$.
\begin{figure}
\centering
\includegraphics[height=5.0cm]{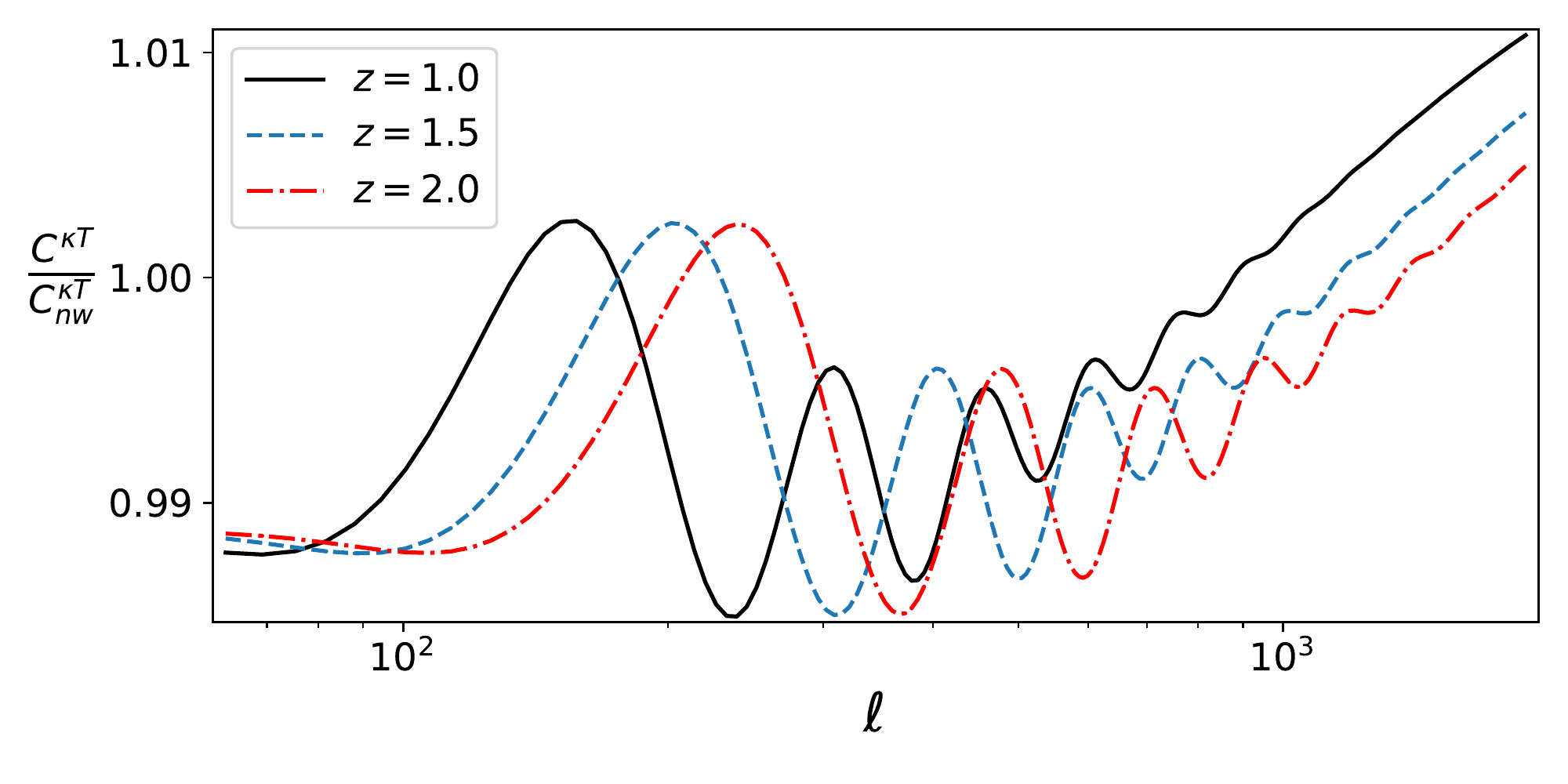}
\caption{This shows the BAO imprint on the transverse cross correlation angular power spectrum $C^{ T \kappa}_\ell$. To highlight the BAO we have divided by the no-wiggles power spectrum $C^{ T \kappa}_{nw}$ which corresponds to the power spectrum without the baryonic feature. This is shown for three
redshifts $z = 1.0, ~1.5, ~2.0 $.}
\label{fig:wiggles}
\end{figure}

\begin{figure}
\centering
\includegraphics[height=3.8cm]{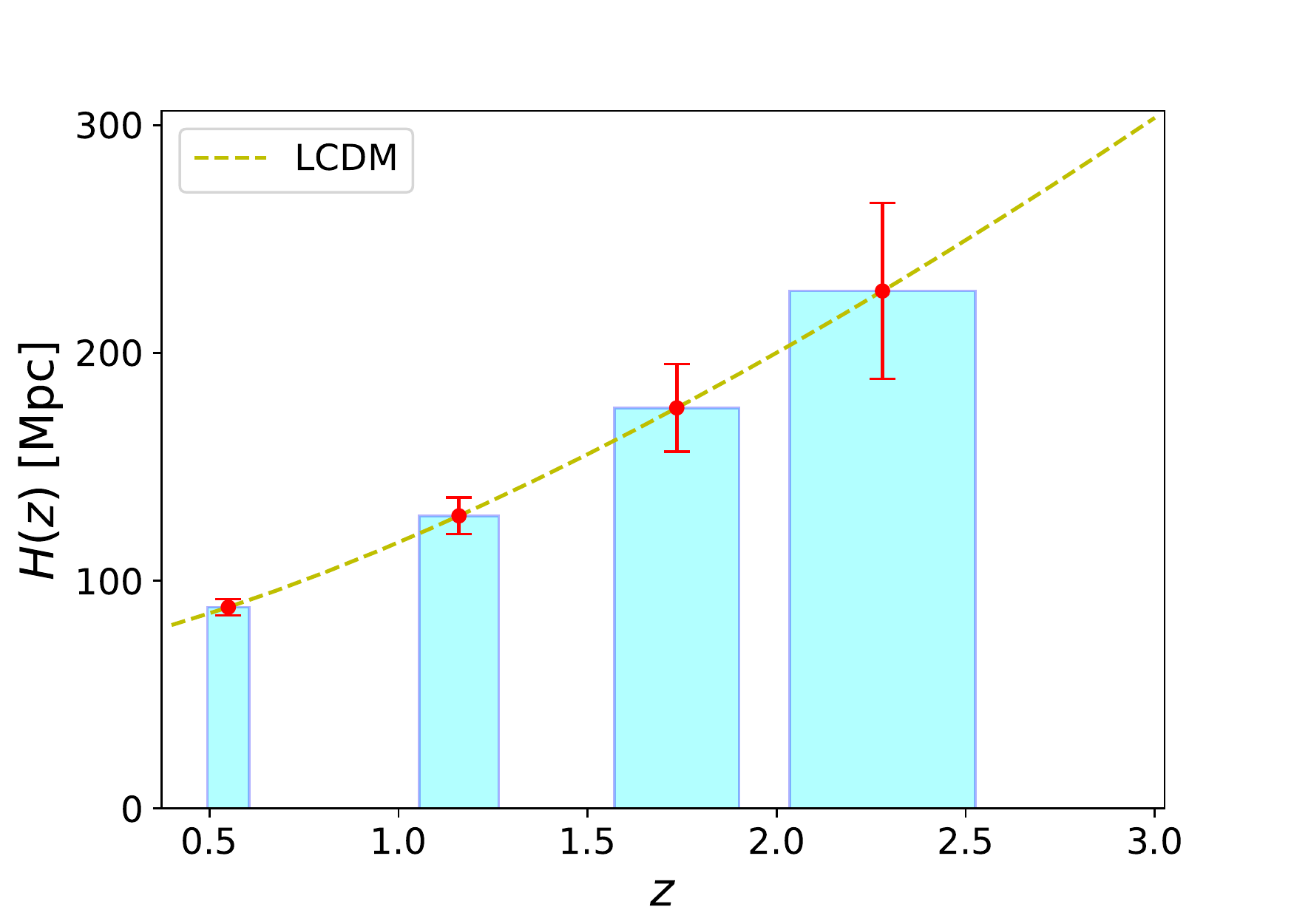}
\includegraphics[height=3.8cm]{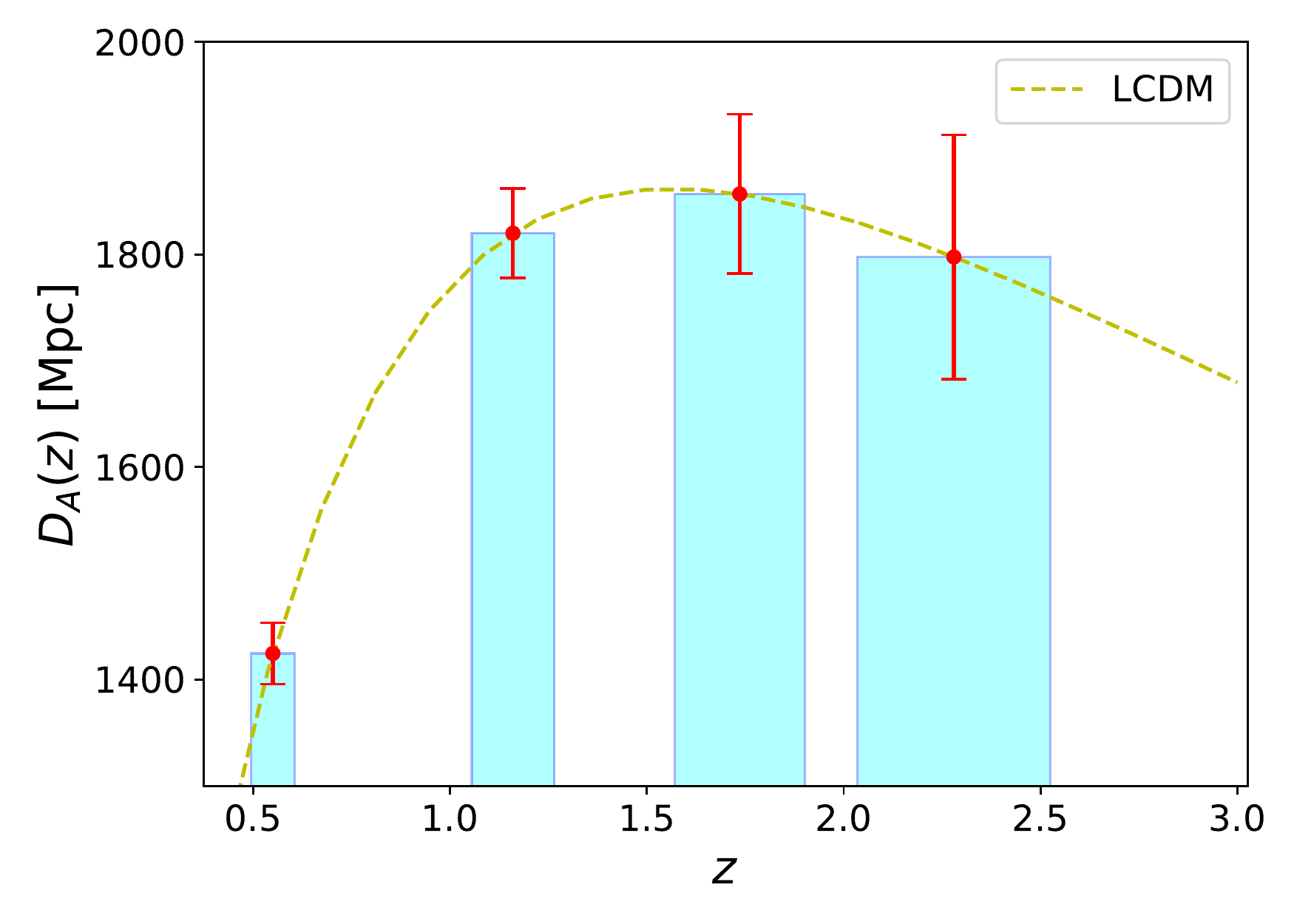} %\par\medskip
\includegraphics[height=3.8cm]{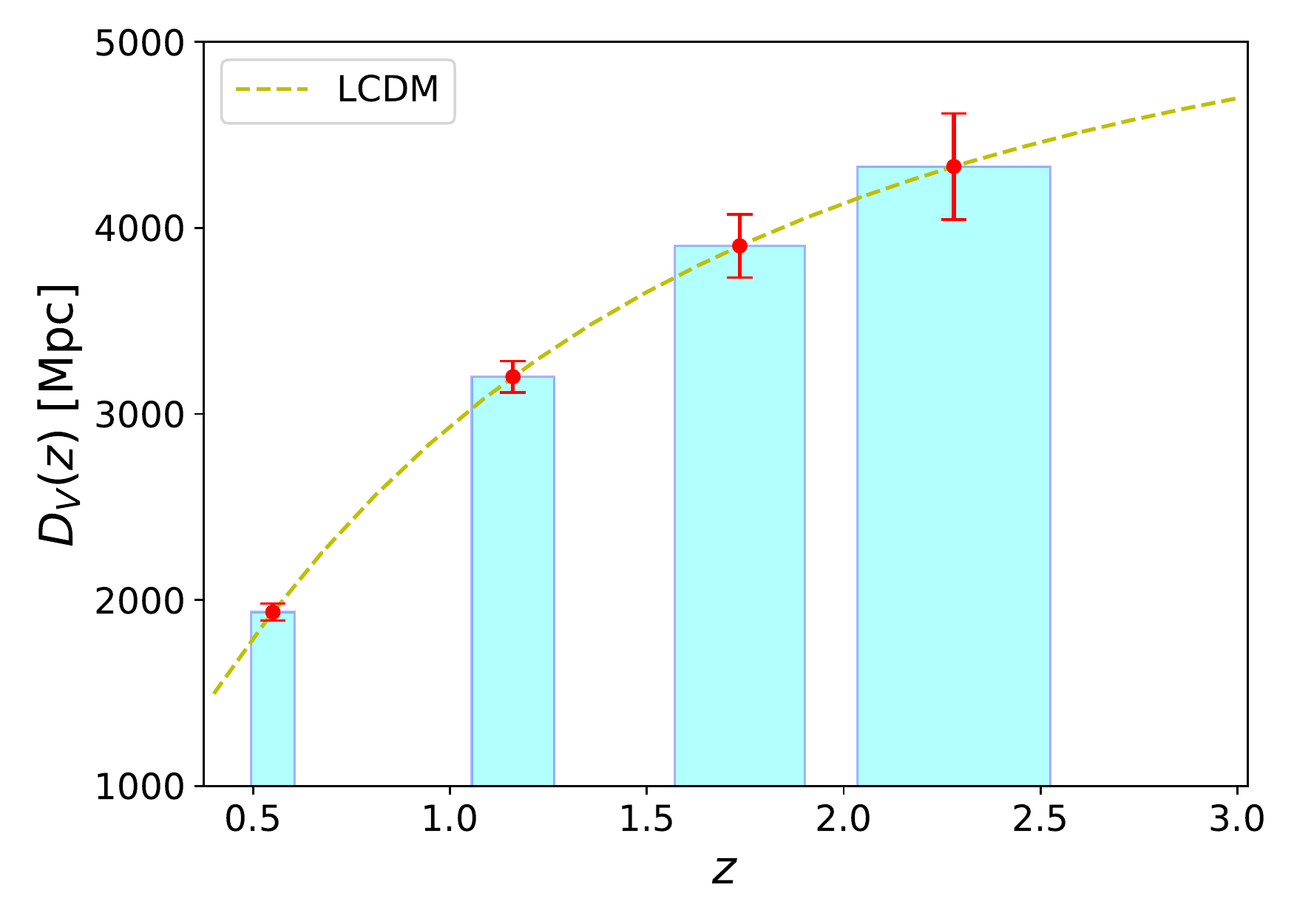}
\caption{The figure shows the projected $1-\sigma$ error bars on $H(z)$, $D_A(z)$ and $D_{v}(z)$ at $4$ redshift bins where the galaxy lensing and HI-21cm cross correlation signal is being observed. The fiducial cosmology is chosen to be LCDM.}
\label{fig:errorestimate}
\end{figure}

\begin{table}
\begin{minipage}{.8\linewidth}
\centering
\begin{tabular}{|c |c | c | c | c |}
\hline \hline
 $N_{ant}$ &Freq. range  & Efficiency & $D_{d}$ & $T_{o}$ \\ [0.5ex] 
 \hline
 $200$ &$400-950$ MHz & 0.7 & $15$m & $600$hrs \\
 \hline
 \hline
  \end{tabular}
  \captionsetup{justification=centering}
\caption{Table showing the parameters of the radio interferometer used for making error projections}
\label{tab:noise-params}
\end{minipage}
\end{table}

\begin{table}
\begin{minipage}{.8\linewidth}
\centering
\begin{tabular}{| c |  c |  c | c |}
\hline \hline
 Redshift$(z)$ & $(\delta H/H) \%$ &$( \delta D_A/D_A) \%$ & $(\delta D_V/D_V) \%$   \\ [0.5ex] 
 \hline \hline
$0.55~$& $4.09~$ &$ 2.02~$ & $2.24~$ \\
 \hline
 $1.16~$& $6.23~$ &$ 2.30~$ & $2.79~$ \\
 \hline
 $1.74$~& $10.90~$ &$ 4.035~$ & $4.62~$ \\
 \hline
 $2.28$& $17.00$ &$ 6.40$ & $6.97$ \\
 \hline
\end{tabular}
  \captionsetup{justification=centering}
\caption{Percentage $1-\sigma$ errors on $D_A$, $H(z)$ and $D_V$.}
\label{tab:errors}
\end{minipage}
\end{table} 
 
The BAO feature  manifests itself as oscillations in the linear matter power spectrum \citep{Hu-eisen}. The first BAO peak  has the largest amplitude and is  a $\sim 10\%$ feature in the matter power spectrum $P(k)$  at $k \approx 0.045 {\rm Mpc}^{-1}$. Figure (\ref{fig:wiggles}) shows the BAO feature in the  cross-correlation angular power spectrum $C^{ T \kappa}_\ell$. The BAO, here, seen projected onto a plane appears
as a series of oscillations in $C^{ T \kappa}_\ell$, The positions of the peaks scales as $\ell \sim k/\chi $.
The amplitude of the first oscillation in $C^{ T \kappa}_\ell$ is the maximum as is about $ 1\%$  in contrast to the $\sim 10 \%$ 
feature seen in $P (k)$. This reduction in amplitude arises due to the projection to a plane
whereby several $3D$ Fourier modes which do not have the BAO feature also contribute to
the $\ell$  where the  BAO peak is seen. For $z = 1.0$ the first peak occurs at $\ell \sim 170$  and it
has a full width of $\Delta \ell \sim 75$. If the redshift is changed,  the position $\ell $  and width  $\Delta \ell$ of the peak both scale as $\chi $.  

We have made error estimates by considering four redshift bins, corresponding to four $32 \rm MHz$ bandwidth radio observations of the 21 cm signal 
at four observing central frequencies. 
The total observing time of $2400$ hrs  is divided into four $600$ hrs observations at each each frequency.

Figure (\ref{fig:errorestimate}) shows the projected errors on $H(z)$ and $D_A (z)$ for the fiducial LCDM cosmology. We find that $D_A(z)$ can be measured at a higher level of precision compared to $D_V(z)$ and $H(z)$. This is because the weak lensing kernel is sensitive to $D_A (z)$ and the integration over $\chi (z)$ in the lensing signal leads to stronger constraints on it. 
The percentage $1-\sigma$ errors are summarized in table (\ref {tab:errors}). We find that $H(z)$ is quite poorly constrained specially at higher redshifts.

\section{Quintessence cosmology}
We investigate spatially flat, homogeneous, and isotropic cosmological models filled with three non-interacting components: dark matter, baryobs and a scalar field $\phi$, minimally coupled with gravity. The Lagrangian density for the quintessence field is given by 
\be
\mathcal{L}_{\phi} = \frac{1}{2} (\partial^\mu \phi \partial_\nu \phi) - V(\phi)
\ee
where $V(\phi)$ is the quintessence potential. The KG equation for quintessence field obtained by varying action w.r.t the $\phi$ is
\be 
\ddot{\phi} + 3H\phi +V_{,\phi} = 0
\ee
where $V_{,\phi}$ differentiation w.r.t $\phi$ and the Friedmann  equation for $H$  is given by
\be 
H^2 = \frac{1}{3}(\rho_m + \rho_b + \rho_\phi)
\ee 
In order to study the dynamics of background quintessence model, let us define the following dimensionless quantities \citep{scherrer2008thawing,amendola_tsujikawa_2010}
%\be
%x = \frac{{\phi'}}{\sqrt{6}M_{pl}}, ~~ y = \frac{\sqrt{V}}{\sqrt{3}HM_{pl}}, ~~ \lambda = -M_{pl} \frac{ V_{,\phi}}{ V} , ~ ~\Gamma = V \frac{V_{,\phi \phi}} { V_{,\phi}^2} , ~~ s = \frac{\sqrt \rho_b}{\sqrt{3}  H} \ee
\be
x = \frac{{\phi'}}{\sqrt{6}}, ~~ y = \frac{\sqrt{V}}{\sqrt{3}H}, ~~ \lambda = - \frac{ V_{,\phi}}{ V} , ~ ~\Gamma = V \frac{V_{,\phi \phi}} { V_{,\phi}^2} , ~~  b  = \frac{\sqrt \rho_b}{\sqrt{3}  H} \ee
where we use units  $ 8\pi G = c = 1$  and  the prime ($'$) denotes the derivative w.r.t the number of e-folding $N = \log(a)$. Using the above quantities we can define the density parameter ($\Omega_\phi$) and the EoS ($w_\phi = p_\phi/\rho_\phi$) to the scalar field as follows
\be
\Omega_{\phi} = x^2 + y^2,~ ~~\gamma = 1+w_{\phi} = \frac{2x^2}{x^2 + y^2}
\ee
The dynamics of background cosmological evolution is obtained by solving a autonomous system of first order equations \citep{scherrer2008thawing,amendola_tsujikawa_2010}. 
\begin{eqnarray}
\gamma ' = 3 \gamma (\gamma - 2 ) + \sqrt{3 \gamma \Omega_\phi}(2 - \gamma)\lambda , \nonumber \\
\Omega'_{\phi}  = 3(1 - \gamma) \Omega_{\phi}(1 -  \Omega_{\phi}), \nonumber \\
\lambda ' = \sqrt{3\gamma \Omega_{\phi}}\lambda ^2 (1 -  \Gamma),\nonumber \\
b'= - \frac{3}{2} b  \Omega_\phi (1- \gamma) 
\label{eq:AutoODE}
\end{eqnarray}
In order to solve the above set of 1st order ODEs numerically, we fix the initial conditions for $\gamma$, $\Omega_{\phi}$, $\lambda$ at the decoupling epoch. For thawing models, the scalar field is initially frozen due to large Hubble damping, and this fixes the initial condition $\gamma_i \approx 0$. The quantity $\Gamma$ which quantifies the shape of the potential is a constant for power law potentials. The parameter $\lambda_i$  is the initial slope of scalar field and measures the deviation of LCDM model. For smaller $\lambda_i$ the EoS ($w_{\phi}$) of scalar field remain close to cosmological constant, whereas larger values of $\lambda_i$ lead to a significant deviation from LCDM. Assuming the contribution of scalar field to the total energy density is negligibly small in the early universe, we fix the present value of $\Omega_{\phi}$.  Similarly, we fix the initial value of $b$ (related to the density parameter for baryons) so that one gets right value of the $\Omega_{b0} = 0.049$ \citep{Planck2018} at the present epoch. 
\begin{figure}
\centering 
\includegraphics[height=5.0cm]{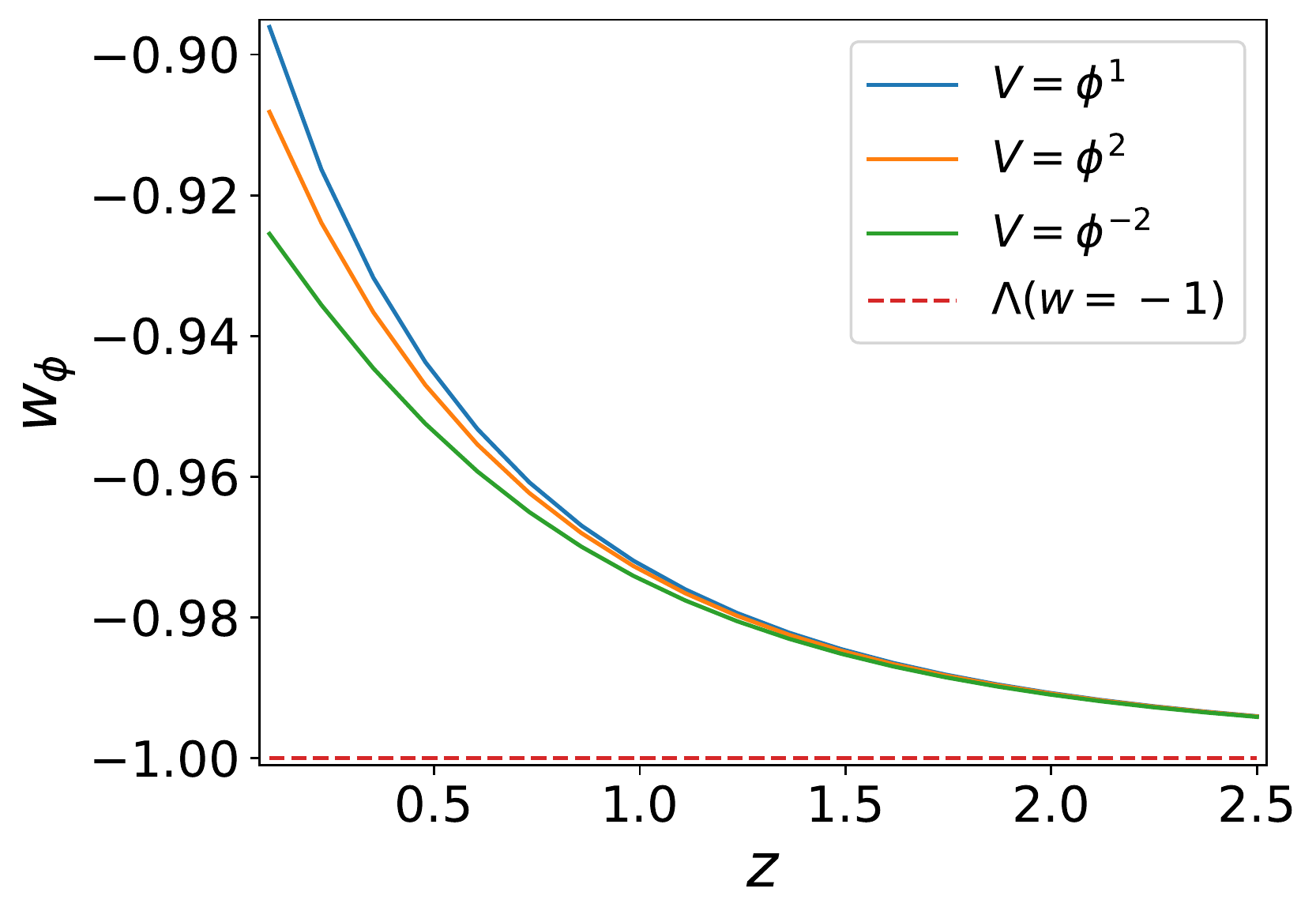}
\caption{The figure shows the EoS ($w_{\phi}$) as a function of redshift z for different quintessence field models after solving the autonomous ODE in (\ref{eq:AutoODE}). We kept the initial slope of the field $\lambda_i = 0.7$ in all the cases.}
\label{fig:wphi}
\end{figure}
Figure (\ref{fig:wphi}) shows the dynamical  evolution of the EoS of quintessence field for three models.
We  note that there is no departure from the LCDM at large redshifts but a prominent model sensitive departure for small redshifts.
At $z\sim 0.5$ there is almost a $\sim 5\%$ departure of the EoS parameter $w_{\phi}$ from that of the non-dynamical cosmological constant. 
The departure of $w_{\phi}$ from its LCDM value of $-1$, imprints on the growing mode of density perturbations by virtue of the changes that it brings to  the Hubble parameter $H(z)$.

Growth of matter fluctuations in the linear regime provides a powerful complementary observation to put tighter constrains on cosmological parameters, and also break the possible degeneracy in diverse dark energy models. We have assumed spatially flat cosmology in our entire analysis and not constrained radiation density, as only  dark matter and dark energy are  dominant in the late universe. The full relativistic treatment of perturbations for Quintessence dark energy has been studied \cite{hussain2016prospects}. Ignoring super-horizon effects,  we note that on sub-horizon scales, ignoring the clustering of Quintessence field, the linearized equations governing the growth of matter fluctuations is given by the ODE \citep{amendola2000coupled,amendola2004linear}
\be
D_{+} ''  + \left( 1 + \frac{\mathcal{H}'(a)}{\mathcal{H}(a)} \right) D_{+}' -  \frac{3}{2} \Omega_{m}(a) D_{+} = 0.
\ee
Here, the prime denotes differentiation w.r.t to \lq $\log a$', $\mathcal{H}$ is the conformal Hubble
parameter defined as $\mathcal{H} = aH$ and $\delta_m$ is the linear density contrast for the dark matter. In order to solve the above ODE, we fix the initial conditions $D_{+} $ grows linearly with $a$ and the first derivative of $\frac{d D_{+} }{d a} =1 $ at early matter dominated epoch ($a= 0.001$). 
We now  consider the BAO imprint on the cross-correlation angular power spectrum to make error predictions on Quintessence dark energy parameters which affects both background evolution and structure formation.
\subsection{Statistical analysis and constraints on model parameters}
We choose the following parameters $( h, \Gamma, \lambda_i, \Omega_{\phi 0})$ to quantify the Quintessence dark energy.
We have use uniform priors for these parameters in the Quintessence model. The Hubble parameter at present (z = 0) in our subsequent calculations is assumed to be $H_0 = 100hKm/s/M pc$, thus define the dimensionless parameter $h$.  We perform  a Markov Chain Monte Carlo (MCMC) analysis using the observational data to constraint the model parameters and evolution of cosmological quantities. The analysis is carried out using the Python  implementation of MCMC sampler introduced by \citet{foreman2013emcee}. We take flat priors for these parameters with ranges of $h \in [0.5,0.9]$, $\Gamma \in [-1.5,1.5]$, $\lambda_i \in [0.5,0.8], \Omega_{\phi 0} \in [0.5,0.8]$ . 

\begin{figure*}
\centering 
\includegraphics[height=8cm]{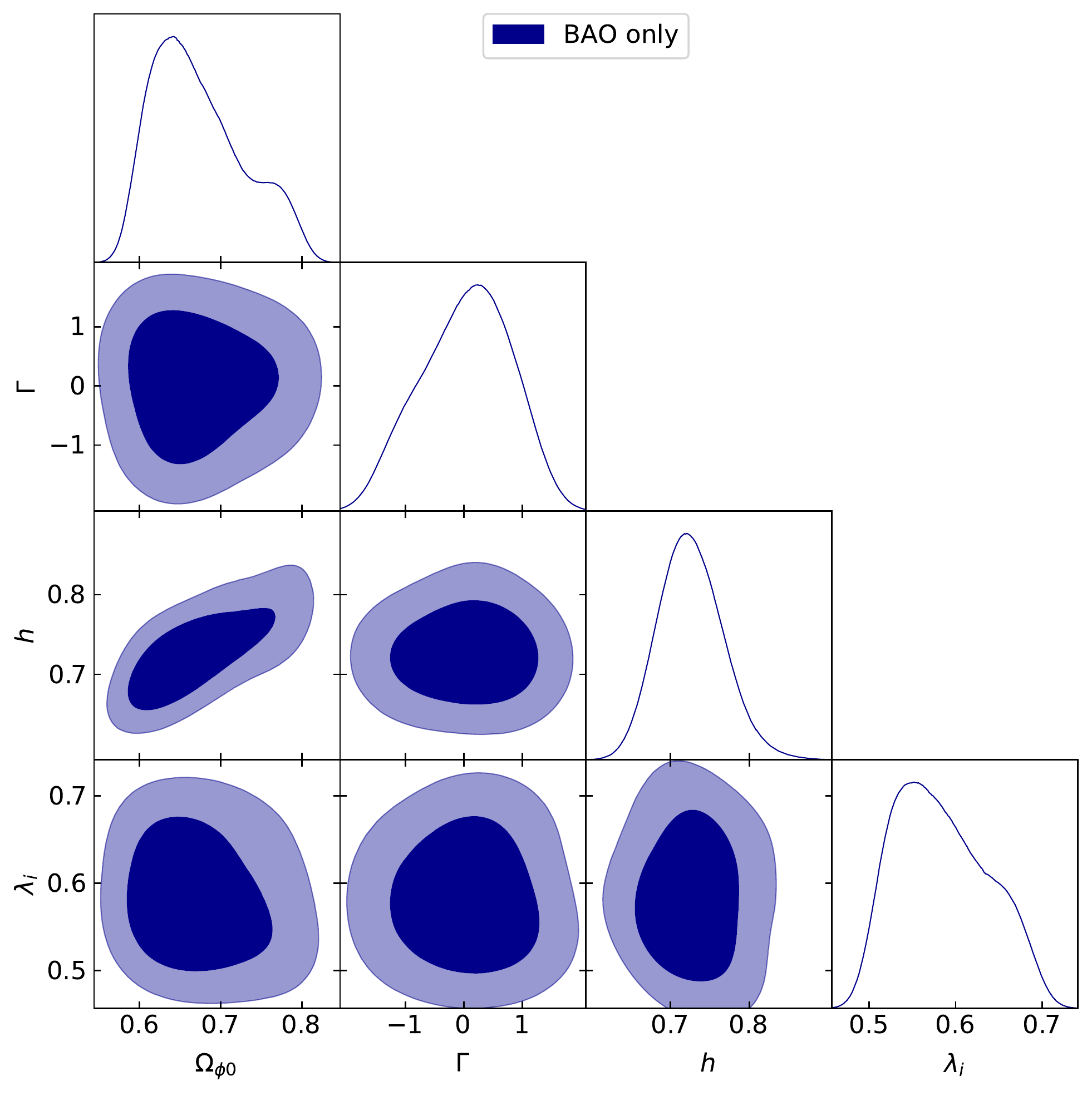}\quad %changeit
\includegraphics[height=8cm]{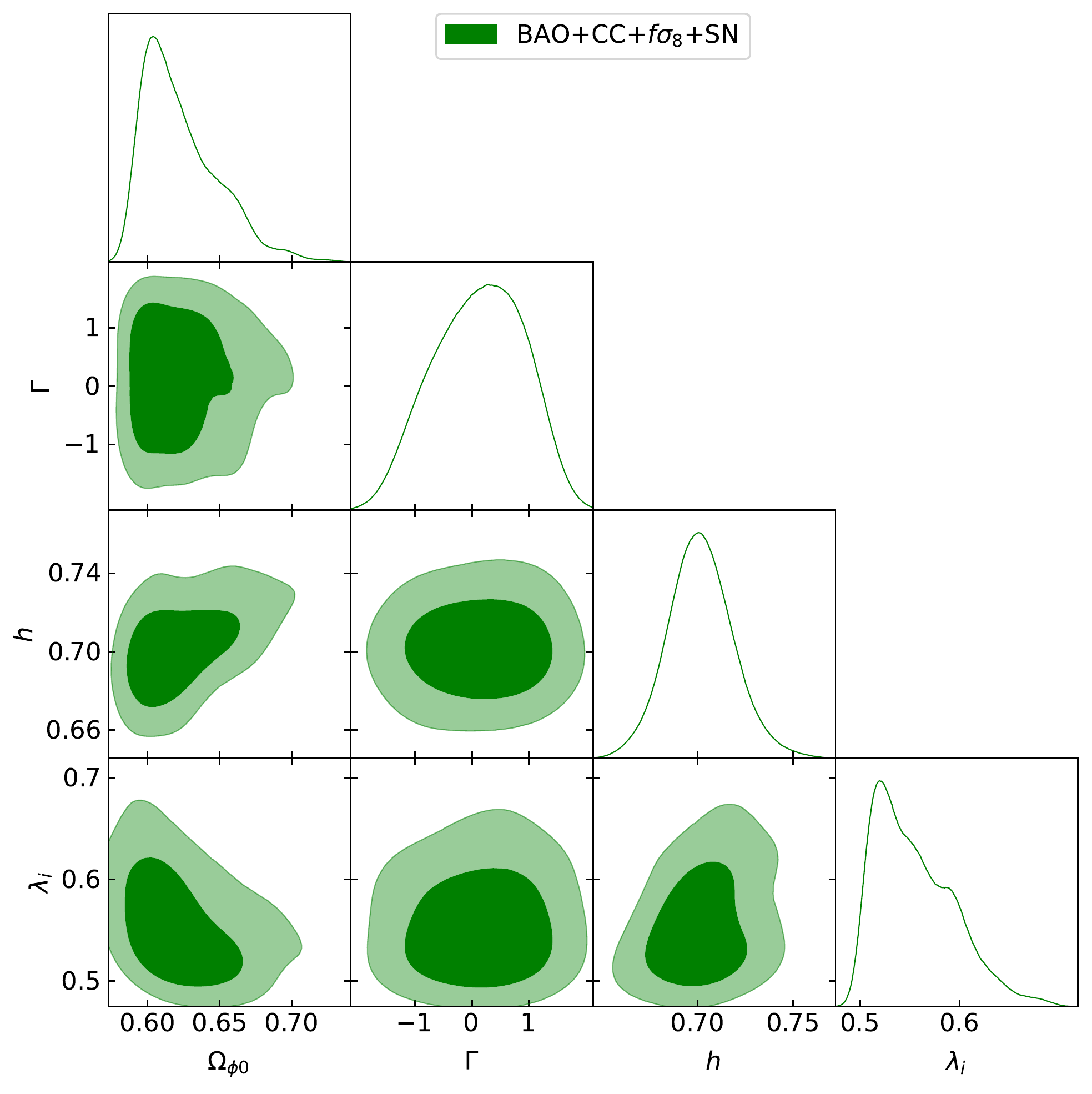}
\caption{Marginalized posterior distribution of the set of parameters  and ($\Omega_{ri},\Omega_{\phi i},\lambda_i, h$) corresponding 2D confidence contours obtained from the MCMC analysis for the model $V(\phi) \sim \phi$. Left panel: utilizing the information from the fisher matrix only. Right panel: utilizing all the data sets mentioned in the discussion on the top of the fisher information.}
\label{fig:mcmc}
\end{figure*}

We first perform the MCMC analysis for the using the error bars obtained on the binned $H(z)$ and $D_A$ from the proposed 21-cm weak lensing cross-correlation.
The figure (\ref{fig:mcmc}) shows the marginalized posterior distribution of the set of parameters  and $( h, \Gamma, \lambda_i, \Omega_{\phi 0})$ the  corresponding 2D confidence contours are obtained for the model $V(\phi) \sim \phi$.  
The results are summarized in table(\ref{tab:MCMC-constraints}).

For a joint  analysis, we employ three mainstream cosmological probes, namely cosmic chronometers (CC), Supernovae Ia (SN) and  $f\sigma_8$. We have used the observational measurements of Hubble expansion rate as a function of redshift using cosmic chronometers (CC) as compiled by \citet{gomez2018h0}. The distance modulus measurement of type Ia supernovae (SN), is adopted from  the Joint Lightcone Analysis sample from \citet{betoule2014improved}. We also incorporated the linear growth rate data, namely the $f\sigma_8(z) (\equiv f(z)\sigma_8 D_m (z))$ from the measurements by various galaxy surveys as compiled by \citet{nesseris2017tension}. 
The posterior probability distributions of the parameters and the corresponding 2D confidence contours are shown in figure (\ref{fig:mcmc}). The constraint obtained for different parameters are shown in table (\ref{tab:MCMC-constraints}). 
The joint analysis gives improved constraints compared to the constraints obtained from the analysis of only our projected BAO results.
These constraints are also competitive with other probes \citep{gupta2012constraining,sangwan2018observational,yang2019constraints}.

\begin{table}
\centering
\begin{tabular}{p{0.15\linewidth}p{0.10\linewidth}p{0.10\linewidth}p{0.10\linewidth}p{0.10\linewidth}}
\hline \hline
 Parameters &$ \Omega_{\phi 0}$ & $ \Gamma$ & $\lambda_i$ & $h$ \\ [0.5ex] 
 \hline\hline
Constraints \\( BAO only) &~~ $0.660^{0.064}_{-0.049} $ & $0.091^{0.784}_{-1.080}$ & $0.575^{0.067}_{-0.050}$ & $0.723^{0.038}_{-0.036}$ \\ 
 \hline
 Constraints \\ (BAO+CC+$f\sigma_8$+SN) & ~~~$0.616^{0.034}_{-0.020} $ & $0.157^{0.895}_{-0.956}$ & $0.548^{0.049}_{-0.036}$ & $0.701^{0.016}_{-0.015}$ \\
 \hline
 \hline
\end{tabular}
\caption{The parameter values, obtained in the MCMC analysis combining all the data sets are tabulated along the $1-\sigma$ uncertainty.}
\label{tab:MCMC-constraints}
\end{table}
\section{Conclusion}
In this paper, we have explored the cross-correlation signal of weak
galaxy lensing and HI 21-cm. From the tomographic study we estimated
the projected errors on the $H(z)$, $D_A (z)$ and $D_V(z)$ over a
redshift range $z \sim 0-3$. The quantities of interest namely $H(z)$
and $D_A(z)$ explicitly appears in the lensing kernel and also in the
BAO feature of the power spectrum.  The cross-angular spectrum involve
a radial integral and hence loses the redshift information. We have
obtained tomographic information by locating the 21-cm slice at
different redshift bins before cross-correlating.

Several observational challenges come in the way of measuring the
cosmological 21-cm signal.  The 21-cm signal is buried deep under
galactic and extra-galactic foregrounds
\citep{2011MNRAS.418.2584G}. We have assumed that this key challenge
is addressed. Even after significant foreground removal, the
cosmological origin of the 21 cm signal can only be ascertained only
through a cross-correlation
\citep{Guha_Sarkar_2010,Carucci_2017,Sarkar_2019}. The foregrounds for
the two individual probes are expected to be significantly
uncorrelated and hence leads to negligible effects in the observing
cross-correlation power spectrum. We have not considered systematic
error which arises from photometric redshift (or so called photo-z)
errors which may significantly degrade the cosmological information in
the context of lensing auto-correlation \citep{Takada_2009}.

The BAO estimates of $H(z)$, $D_A (z)$ allows us to probe dark energy
models. We have considered the quintessence scalar field as a
potential dark energy candidate and studied the background dynamics as
well as the growth perturbation in linear regime in such a paradigm.
A Baysean parameter estimation using our BAO estimates indicate the
possibility of good constraints on scalar field models. The
constraints also improve when joint analysis with other probes is
undertaken and reaches precision levels competitive with the existing
literature.

%%%%%%%%%%%%%%%%%%%%%%%%%%%%
\bibliographystyle{mn2e}
\bibliography{references}

\begin{thebibliography}{}

\bibitem[\protect\citeauthoryear{Abbott, Abdalla, Aleksi{\'c}, Allam, Amara,
  Bacon, Balbinot, Banerji, Bechtol, Benoit-L{\'e}vy et~al.,}{Abbott
  et~al.}{2016}]{abbott2016dark}
Abbott T.,  Abdalla F.~B.,  Aleksi{\'c} J.,  Allam S.,  Amara A.,  Bacon D.,
  Balbinot E.,  Banerji M.,  Bechtol K.,  Benoit-L{\'e}vy A.,    et~al., 2016,
  Monthly Notices of the Royal Astronomical Society, 460, 1270

\bibitem[\protect\citeauthoryear{Aghanim, Akrami, Ashdown, Aumont, Baccigalupi,
  Ballardini, Banday, Barreiro, Bartolo \& et al.}{Aghanim
  et~al.}{2020}]{Planck2018}
Aghanim N.,  Akrami Y.,  Ashdown M.,  Aumont J.,  Baccigalupi C.,  Ballardini
  M.,  Banday A.~J.,  Barreiro R.~B.,  Bartolo N.,    et al. 2020, Astronomy
  and Astrophysics, 641, 6

\bibitem[\protect\citeauthoryear{Aihara, Arimoto, Armstrong, Arnouts, Bahcall,
  Bickerton, Bosch, Bundy, Capak, Chan et~al.,}{Aihara
  et~al.}{2018}]{aihara2018hyper}
Aihara H.,  Arimoto N.,  Armstrong R.,  Arnouts S.,  Bahcall N.~A.,  Bickerton
  S.,  Bosch J.,  Bundy K.,  Capak P.~L.,  Chan J.~H.,    et~al., 2018,
  Publications of the Astronomical Society of Japan, 70, S4

\bibitem[\protect\citeauthoryear{Amendola}{Amendola}{2000}]{amendola2000coupled}
Amendola L.,  2000, Physical Review D, 62, 043511

\bibitem[\protect\citeauthoryear{Amendola}{Amendola}{2004}]{amendola2004linear}
Amendola L.,  2004, Physical Review D, 69, 103524

\bibitem[\protect\citeauthoryear{Amendola \& Tsujikawa}{Amendola \&
  Tsujikawa}{2010}]{amendola_tsujikawa_2010}
Amendola L.,  Tsujikawa S.,  2010, Dark Energy: Theory and Observations.
Cambridge University Press

\bibitem[\protect\citeauthoryear{Anderson, Aubourg \& et.al}{Anderson
  et~al.}{2012}]{Anderson_2012}
Anderson L.,  Aubourg E.,    et.al B.,  2012, Monthly Notices of the Royal
  Astronomical Society, 427, 3435

\bibitem[\protect\citeauthoryear{Bagla, Khandai \& Datta}{Bagla
  et~al.}{2010}]{Bagla_2010}
Bagla J.~S.,  Khandai N.,    Datta K.~K.,  2010, Monthly Notices of the Royal
  Astronomical Society, 407, 567–580

\bibitem[\protect\citeauthoryear{Bamba, Capozziello, Nojiri \& Odintsov}{Bamba
  et~al.}{2012}]{bamba2012dark}
Bamba K.,  Capozziello S.,  Nojiri S.,    Odintsov S.~D.,  2012, Astrophysics
  and Space Science, 342, 155

\bibitem[\protect\citeauthoryear{Bartelmann \& Schneider}{Bartelmann \&
  Schneider}{2001}]{Bartelmann_2001}
Bartelmann M.,  Schneider P.,  2001, Physics Reports, 340, 291–472

\bibitem[\protect\citeauthoryear{Betoule, Kessler, Guy, Mosher, Hardin, Biswas,
  Astier, El-Hage, Konig, Kuhlmann et~al.,}{Betoule
  et~al.}{2014}]{betoule2014improved}
Betoule M.,  Kessler R.,  Guy J.,  Mosher J.,  Hardin D.,  Biswas R.,  Astier
  P.,  El-Hage P.,  Konig M.,  Kuhlmann S.,    et~al., 2014, Astronomy \&
  Astrophysics, 568, A22

\bibitem[\protect\citeauthoryear{{Bharadwaj} \& {Ali}}{{Bharadwaj} \&
  {Ali}}{2005}]{bali}
{Bharadwaj} S.,  {Ali} S.~S.,  2005, MNRAS, 356, 1519

\bibitem[\protect\citeauthoryear{{Bharadwaj}, {Nath} \& {Sethi}}{{Bharadwaj}
  et~al.}{2001}]{poreion2}
{Bharadwaj} S.,  {Nath} B.~B.,    {Sethi} S.~K.,  2001, Journal of Astrophysics
  and Astronomy, 22, 21

\bibitem[\protect\citeauthoryear{{Bharadwaj} \& {Pandey}}{{Bharadwaj} \&
  {Pandey}}{2003}]{poreion7}
{Bharadwaj} S.,  {Pandey} S.~K.,  2003, Journal of Astrophysics and Astronomy,
  24, 23

\bibitem[\protect\citeauthoryear{{Bharadwaj} \& {Sethi}}{{Bharadwaj} \&
  {Sethi}}{2001}]{poreion1}
{Bharadwaj} S.,  {Sethi} S.~K.,  2001, Journal of Astrophysics and Astronomy,
  22, 293

\bibitem[\protect\citeauthoryear{{Bharadwaj}, {Sethi} \& {Saini}}{{Bharadwaj}
  et~al.}{2009}]{param3}
{Bharadwaj} S.,  {Sethi} S.~K.,    {Saini} T.~D.,  2009, Physical Rev D, 79,
  083538

\bibitem[\protect\citeauthoryear{{Bharadwaj} \& {Srikant}}{{Bharadwaj} \&
  {Srikant}}{2004}]{poreion8}
{Bharadwaj} S.,  {Srikant} P.~S.,  2004, Journal of Astrophysics and Astronomy,
  25, 67

\bibitem[\protect\citeauthoryear{Caldwell, Dave \& Steinhardt}{Caldwell
  et~al.}{1998}]{Steinhardt_1998}
Caldwell R.~R.,  Dave R.,    Steinhardt P.~J.,  1998, Phys. Rev. Lett., 80,
  1582

\bibitem[\protect\citeauthoryear{Carucci, Villaescusa-Navarro \& Viel}{Carucci
  et~al.}{2017}]{Carucci_2017}
Carucci I.~P.,  Villaescusa-Navarro F.,    Viel M.,  2017, Journal of Cosmology
  and Astroparticle Physics, 2017, 001–001

\bibitem[\protect\citeauthoryear{Chang, Jarvis, Jain, Kahn, Kirkby, Connolly,
  Krughoff, Peng \& Peterson}{Chang et~al.}{2013}]{chang2013effective}
Chang C.,  Jarvis M.,  Jain B.,  Kahn S.,  Kirkby D.,  Connolly A.,  Krughoff
  S.,  Peng E.-H.,    Peterson J.,  2013, Monthly Notices of the Royal
  Astronomical Society, 434, 2121

\bibitem[\protect\citeauthoryear{{Chang}, {Pen}, {Peterson} \&
  {McDonald}}{{Chang} et~al.}{2008}]{param2}
{Chang} T.,  {Pen} U.,  {Peterson} J.~B.,    {McDonald} P.,  2008, Physical
  Review Letters, 100, 091303

\bibitem[\protect\citeauthoryear{Copeland, Sami \& Tsujikawa}{Copeland
  et~al.}{2006}]{copeland2006dynamics}
Copeland E.~J.,  Sami M.,    Tsujikawa S.,  2006, International Journal of
  Modern Physics D, 15, 1753

\bibitem[\protect\citeauthoryear{Dash \& Guha~Sarkar}{Dash \&
  Guha~Sarkar}{2021}]{Dash_2021}
Dash C.~B.,  Guha~Sarkar T.,  2021, Journal of Cosmology and Astroparticle
  Physics, 2021, 016

\bibitem[\protect\citeauthoryear{Datta, Choudhury \& Bharadwaj}{Datta
  et~al.}{2007}]{datta2007multifrequency}
Datta K.~K.,  Choudhury T.~R.,    Bharadwaj S.,  2007, Monthly Notices of the
  Royal Astronomical Society, 378, 119

\bibitem[\protect\citeauthoryear{Eisenstein \& Hu}{Eisenstein \&
  Hu}{1998}]{Hu-eisen}
Eisenstein D.~J.,  Hu W.,  1998, The Astrophysical Journal, 496, 605

\bibitem[\protect\citeauthoryear{Eisenstein, Zehavi, Hogg, Scoccimarro,
  Blanton, Nichol, Scranton, Seo, Tegmark, Zheng \& et al.}{Eisenstein
  et~al.}{2005}]{Eisenstein_2005}
Eisenstein D.~J.,  Zehavi I.,  Hogg D.~W.,  Scoccimarro R.,  Blanton M.~R.,
  Nichol R.~C.,  Scranton R.,  Seo H.,  Tegmark M.,  Zheng Z.,    et al. 2005,
  The Astrophysical Journal, 633, 560–574

\bibitem[\protect\citeauthoryear{Fang, Bi, Xiang \& Boerner}{Fang
  et~al.}{1993}]{fang}
Fang L.~Z.,  Bi H.,  Xiang S.,    Boerner G.,  1993, The Astrophysical Journal,
  413, 477

\bibitem[\protect\citeauthoryear{Fonseca, Maartens \& Santos}{Fonseca
  et~al.}{2017}]{fonseca2017probing}
Fonseca J.,  Maartens R.,    Santos M.~G.,  2017, Monthly Notices of the Royal
  Astronomical Society, 466, 2780

\bibitem[\protect\citeauthoryear{Foreman-Mackey, Hogg, Lang \&
  Goodman}{Foreman-Mackey et~al.}{2013}]{foreman2013emcee}
Foreman-Mackey D.,  Hogg D.~W.,  Lang D.,    Goodman J.,  2013, Publications of
  the Astronomical Society of the Pacific, 125, 306

\bibitem[\protect\citeauthoryear{Gallerani, Choudhury \& Ferrara}{Gallerani
  et~al.}{2006}]{Gallerani_2006}
Gallerani S.,  Choudhury T.~R.,    Ferrara A.,  2006, Monthly Notices of the
  Royal Astronomical Society, 370, 1401–1421

\bibitem[\protect\citeauthoryear{Geil, Gaensler \& Wyithe}{Geil
  et~al.}{2011}]{geil2011polarized}
Geil P.~M.,  Gaensler B.,    Wyithe J. S.~B.,  2011, Monthly Notices of the
  Royal Astronomical Society, 418, 516

\bibitem[\protect\citeauthoryear{{Ghosh}, {Bharadwaj}, {Ali} \&
  {Chengalur}}{{Ghosh} et~al.}{2011}]{2011MNRAS.418.2584G}
{Ghosh} A.,  {Bharadwaj} S.,  {Ali} S.~S.,    {Chengalur} J.~N.,  2011, MNRAS,
  418, 2584

\bibitem[\protect\citeauthoryear{G{\'o}mez-Valent \& Amendola}{G{\'o}mez-Valent
  \& Amendola}{2018}]{gomez2018h0}
G{\'o}mez-Valent A.,  Amendola L.,  2018, Journal of Cosmology and
  Astroparticle Physics, 2018, 051

\bibitem[\protect\citeauthoryear{Guha~Sarkar, Bharadwaj, Choudhury \&
  Datta}{Guha~Sarkar et~al.}{2010}]{Guha_Sarkar_2010}
Guha~Sarkar T.,  Bharadwaj S.,  Choudhury T.~R.,    Datta K.~K.,  2010, Monthly
  Notices of the Royal Astronomical Society, 410, 1130–1134

\bibitem[\protect\citeauthoryear{Guha~Sarkar, Mitra, Majumdar \&
  Choudhury}{Guha~Sarkar et~al.}{2012}]{Guha_Sarkar_2012}
Guha~Sarkar T.,  Mitra S.,  Majumdar S.,    Choudhury T.~R.,  2012, Monthly
  Notices of the Royal Astronomical Society, 421, 3570–3578

\bibitem[\protect\citeauthoryear{Gupta, Majumdar \& Sen}{Gupta
  et~al.}{2012}]{gupta2012constraining}
Gupta G.,  Majumdar S.,    Sen A.~A.,  2012, Monthly Notices of the Royal
  Astronomical Society, 420, 1309

\bibitem[\protect\citeauthoryear{Hu}{Hu}{1999}]{Hu_1999}
Hu W.,  1999, The Astrophysical Journal, 522, L21–L24

\bibitem[\protect\citeauthoryear{Hu \& Sawicki}{Hu \& Sawicki}{2007}]{Hu_2007}
Hu W.,  Sawicki I.,  2007, Physical Review D, 76

\bibitem[\protect\citeauthoryear{Hu \& Sugiyama}{Hu \&
  Sugiyama}{1996}]{hu1996small}
Hu W.,  Sugiyama N.,  1996, The Astrophysical Journal, 471, 542

\bibitem[\protect\citeauthoryear{Hussain, Thakur, Guha~Sarkar \& Sen}{Hussain
  et~al.}{2016}]{hussain2016prospects}
Hussain A.,  Thakur S.,  Guha~Sarkar T.,    Sen A.~A.,  2016, Monthly Notices
  of the Royal Astronomical Society, 463, 3492

\bibitem[\protect\citeauthoryear{Ivezi{\'c}, Axelrod, Brandt, Burke, Claver,
  Connolly, Cook, Gee, Gilmore, Jacoby et~al.,}{Ivezi{\'c}
  et~al.}{2008}]{ivezic2008large}
Ivezi{\'c} {\v{Z}}.,  Axelrod T.,  Brandt W.,  Burke D.,  Claver C.,  Connolly
  A.,  Cook K.,  Gee P.,  Gilmore D.,  Jacoby S.,    et~al., 2008, Serbian
  Astronomical Journal, pp 1--13

\bibitem[\protect\citeauthoryear{Khoury \& Weltman}{Khoury \&
  Weltman}{2004}]{Khoury_2004}
Khoury J.,  Weltman A.,  2004, Physical Review D, 69

\bibitem[\protect\citeauthoryear{Komatsu, Dunkley, Nolta, Bennett, Gold,
  Hinshaw, Jarosik, Larson, Limon, Page \& et al.}{Komatsu
  et~al.}{2009}]{Komatsu_2009}
Komatsu E.,  Dunkley J.,  Nolta M.~R.,  Bennett C.~L.,  Gold B.,  Hinshaw G.,
  Jarosik N.,  Larson D.,  Limon M.,  Page L.,    et al. 2009, The
  Astrophysical Journal Supplement Series, 180, 330–376

\bibitem[\protect\citeauthoryear{Laureijs, Amiaux, Arduini, Augueres,
  Brinchmann, Cole, Cropper, Dabin, Duvet, Ealet et~al.,}{Laureijs
  et~al.}{2011}]{laureijs2011euclid}
Laureijs R.,  Amiaux J.,  Arduini S.,  Augueres J.-L.,  Brinchmann J.,  Cole
  R.,  Cropper M.,  Dabin C.,  Duvet L.,  Ealet A.,    et~al., 2011, arXiv
  preprint arXiv:1110.3193

\bibitem[\protect\citeauthoryear{{Loeb} \& {Wyithe}}{{Loeb} \&
  {Wyithe}}{2008}]{poreion4}
{Loeb} A.,  {Wyithe} J.~S.~B.,  2008, Physical Review Letters, 100, 161301

\bibitem[\protect\citeauthoryear{Mao, Tegmark, McQuinn, Zaldarriaga \&
  Zahn}{Mao et~al.}{2008}]{Mao_2008}
Mao Y.,  Tegmark M.,  McQuinn M.,  Zaldarriaga M.,    Zahn O.,  2008, Physical
  Review D, 78

\bibitem[\protect\citeauthoryear{{Mao}, {Tegmark}, {McQuinn}, {Zaldarriaga} \&
  {Zahn}}{{Mao} et~al.}{2008}]{param4}
{Mao} Y.,  {Tegmark} M.,  {McQuinn} M.,  {Zaldarriaga} M.,    {Zahn} O.,  2008,
  Physical Rev D, 78, 023529

\bibitem[\protect\citeauthoryear{Marín, Gnedin, Seo \& Vallinotto}{Marín
  et~al.}{2010}]{Mar_n_2010}
Marín F.~A.,  Gnedin N.~Y.,  Seo H.-J.,    Vallinotto A.,  2010, The
  Astrophysical Journal, 718, 972–980

\bibitem[\protect\citeauthoryear{Nesseris, Pantazis \&
  Perivolaropoulos}{Nesseris et~al.}{2017}]{nesseris2017tension}
Nesseris S.,  Pantazis G.,    Perivolaropoulos L.,  2017, Physical Review D,
  96, 023542

\bibitem[\protect\citeauthoryear{Nojiri \& Odintsov}{Nojiri \&
  Odintsov}{2007}]{nojiri2007introduction}
Nojiri S.,  Odintsov S.~D.,  2007, International Journal of Geometric Methods
  in Modern Physics, 4, 115

\bibitem[\protect\citeauthoryear{Noterdaeme, Petitjean, Ledoux \&
  Srianand}{Noterdaeme et~al.}{2009}]{Noterdaeme_2009}
Noterdaeme P.,  Petitjean P.,  Ledoux C.,    Srianand R.,  2009, Astronomy and
  Astrophysics, 505, 1087–1098

\bibitem[\protect\citeauthoryear{{Padmanabhan}, Choudhury \&
  Refregier}{{Padmanabhan} et~al.}{2015}]{poreion12}
{Padmanabhan} H.,  Choudhury T.~R.,    Refregier A.,  2015, Monthly Notices of
  the Royal Astronomical Society, 447, 3745

\bibitem[\protect\citeauthoryear{Panda, Sumitomo \& Trivedi}{Panda
  et~al.}{2011}]{panda2011axions}
Panda S.,  Sumitomo Y.,    Trivedi S.~P.,  2011, Physical Review D, 83, 083506

\bibitem[\protect\citeauthoryear{Peebles \& Ratra}{Peebles \&
  Ratra}{2003}]{RevModPhys.75.559}
Peebles P. J.~E.,  Ratra B.,  2003, Rev. Mod. Phys., 75, 559

\bibitem[\protect\citeauthoryear{Percival, Cole, Eisenstein, Nichol, Peacock,
  Pope \& Szalay}{Percival et~al.}{2007}]{Percival_2007}
Percival W.~J.,  Cole S.,  Eisenstein D.~J.,  Nichol R.~C.,  Peacock J.~A.,
  Pope A.~C.,    Szalay A.~S.,  2007, Monthly Notices of the Royal Astronomical
  Society, 381, 1053–1066

\bibitem[\protect\citeauthoryear{Perlmutter, Gabi, Goldhaber, Goobar, Groom,
  Hook, Kim, Kim, Lee \& Pain}{Perlmutter et~al.}{1997}]{Perlmutter_1997}
Perlmutter S.,  Gabi S.,  Goldhaber G.,  Goobar A.,  Groom D.~E.,  Hook I.~M.,
  Kim A.~G.,  Kim M.~Y.,  Lee J.~C.,    Pain R. e.~a.,  1997, The Astrophysical
  Journal, 483, 565–581

\bibitem[\protect\citeauthoryear{Ratra \& Peebles}{Ratra \&
  Peebles}{1988}]{Ratra-Peebles_1988}
Ratra B.,  Peebles P. J.~E.,  1988, Phys. Rev. D, 37, 3406

\bibitem[\protect\citeauthoryear{Riess, Filippenko, Challis, Clocchiatti,
  Diercks, Garnavich, Gilliland, Hogan, Jha, Kirshner et~al.,}{Riess
  et~al.}{1998}]{riess1998observational}
Riess A.~G.,  Filippenko A.~V.,  Challis P.,  Clocchiatti A.,  Diercks A.,
  Garnavich P.~M.,  Gilliland R.~L.,  Hogan C.~J.,  Jha S.,  Kirshner R.~P.,
  et~al., 1998, The Astronomical Journal, 116, 1009

\bibitem[\protect\citeauthoryear{Riess, Macri, Hoffmann, Scolnic, Casertano,
  Filippenko, Tucker, Reid, Jones, Silverman \& et al.}{Riess
  et~al.}{2016}]{Riess_2016}
Riess A.~G.,  Macri L.~M.,  Hoffmann S.~L.,  Scolnic D.,  Casertano S.,
  Filippenko A.~V.,  Tucker B.~E.,  Reid M.~J.,  Jones D.~O.,  Silverman J.~M.,
     et al. 2016, The Astrophysical Journal, 826, 56

\bibitem[\protect\citeauthoryear{Sahni \& Starobinsky}{Sahni \&
  Starobinsky}{2000}]{sahni2000case}
Sahni V.,  Starobinsky A.,  2000, International Journal of Modern Physics D, 9,
  373

\bibitem[\protect\citeauthoryear{Sangwan, Tripathi \& Jassal}{Sangwan
  et~al.}{2018}]{sangwan2018observational}
Sangwan A.,  Tripathi A.,    Jassal H.,  2018, arXiv preprint arXiv:1804.09350

\bibitem[\protect\citeauthoryear{Sarkar, Pal \& Sarkar}{Sarkar
  et~al.}{2019}]{Sarkar_2019}
Sarkar A.,  Pal A.~K.,    Sarkar T.~G.,  2019, Journal of Cosmology and
  Astroparticle Physics, 2019, 058–058

\bibitem[\protect\citeauthoryear{Sarkar, Bharadwaj \& Anathpindika}{Sarkar
  et~al.}{2016}]{Sarkar_2016}
Sarkar D.,  Bharadwaj S.,    Anathpindika S.,  2016, Monthly Notices of the
  Royal Astronomical Society, 460, 4310–4319

\bibitem[\protect\citeauthoryear{Sarkar}{Sarkar}{2010}]{GSarkar_2010}
Sarkar T.~G.,  2010, Journal of Cosmology and Astroparticle Physics, 2010,
  002–002

\bibitem[\protect\citeauthoryear{Sarkar \& Bharadwaj}{Sarkar \&
  Bharadwaj}{2011}]{sarkar2011imprint}
Sarkar T.~G.,  Bharadwaj S.,  2011, arXiv preprint arXiv:1112.0745

\bibitem[\protect\citeauthoryear{Sarkar \& Bharadwaj}{Sarkar \&
  Bharadwaj}{2013}]{sarkar2013predictions}
Sarkar T.~G.,  Bharadwaj S.,  2013, Journal of Cosmology and Astroparticle
  Physics, 2013, 023

\bibitem[\protect\citeauthoryear{Sarkar \& Datta}{Sarkar \&
  Datta}{2015}]{Sarkar_2015}
Sarkar T.~G.,  Datta K.~K.,  2015, Journal of Cosmology and Astroparticle
  Physics, 2015, 001–001

\bibitem[\protect\citeauthoryear{Sarkar, Datta \& Bharadwaj}{Sarkar
  et~al.}{2009}]{Sarkar_2009}
Sarkar T.~G.,  Datta K.~K.,    Bharadwaj S.,  2009, Journal of Cosmology and
  Astroparticle Physics, 2009, 019–019

\bibitem[\protect\citeauthoryear{Scherrer \& Sen}{Scherrer \&
  Sen}{2008}]{scherrer2008thawing}
Scherrer R.~J.,  Sen A.,  2008, Physical Review D, 77, 083515

\bibitem[\protect\citeauthoryear{Seo \& Eisenstein}{Seo \&
  Eisenstein}{2007}]{seo2007improved}
Seo H.-J.,  Eisenstein D.~J.,  2007, The Astrophysical Journal, 665, 14

\bibitem[\protect\citeauthoryear{Shoji, Jeong \& Komatsu}{Shoji
  et~al.}{2009}]{shoji2009extracting}
Shoji M.,  Jeong D.,    Komatsu E.,  2009, The Astrophysical Journal, 693, 1404

\bibitem[\protect\citeauthoryear{Spergel, Gehrels, Baltay, Bennett,
  Breckinridge, Donahue, Dressler, Gaudi, Greene, Guyon et~al.,}{Spergel
  et~al.}{2015}]{spergel2015wide}
Spergel D.,  Gehrels N.,  Baltay C.,  Bennett D.,  Breckinridge J.,  Donahue
  M.,  Dressler A.,  Gaudi B.,  Greene T.,  Guyon O.,    et~al., 2015, arXiv
  preprint arXiv:1503.03757

\bibitem[\protect\citeauthoryear{Starobinsky}{Starobinsky}{2007}]{Starobinsky_2007}
Starobinsky A.~A.,  2007, JETP Letters, 86, 157–163

\bibitem[\protect\citeauthoryear{Steinhardt, Wang \& Zlatev}{Steinhardt
  et~al.}{1999}]{PhysRevD.59.123504}
Steinhardt P.~J.,  Wang L.,    Zlatev I.,  1999, Phys. Rev. D, 59, 123504

\bibitem[\protect\citeauthoryear{Takada \& Jain}{Takada \&
  Jain}{2004}]{takada2004cosmological}
Takada M.,  Jain B.,  2004, Monthly Notices of the Royal Astronomical Society,
  348, 897

\bibitem[\protect\citeauthoryear{Takada \& Jain}{Takada \&
  Jain}{2009}]{Takada_2009}
Takada M.,  Jain B.,  2009, Monthly Notices of the Royal Astronomical Society,
  395, 2065–2086

\bibitem[\protect\citeauthoryear{Vallinotto, Das, Spergel \& Viel}{Vallinotto
  et~al.}{2009}]{Vallinotto_2009}
Vallinotto A.,  Das S.,  Spergel D.~N.,    Viel M.,  2009, Physical Review
  Letters, 103

\bibitem[\protect\citeauthoryear{Villaescusa-Navarro, Viel, Datta \&
  Choudhury}{Villaescusa-Navarro et~al.}{2014}]{villaescusa2014modeling}
Villaescusa-Navarro F.,  Viel M.,  Datta K.~K.,    Choudhury T.~R.,  2014,
  Journal of Cosmology and Astroparticle Physics, 2014, 050

\bibitem[\protect\citeauthoryear{{Visbal}, {Loeb} \& {Wyithe}}{{Visbal}
  et~al.}{2009}]{poreion6}
{Visbal} E.,  {Loeb} A.,    {Wyithe} S.,  2009, Journal of Cosmology and
  Astro-Particle Physics, 10, 30

\bibitem[\protect\citeauthoryear{Waerbeke \& Mellier}{Waerbeke \&
  Mellier}{2003}]{waerbeke2003}
Waerbeke L.~V.,  Mellier Y., , 2003, Gravitational Lensing by Large Scale
  Structures: A Review

\bibitem[\protect\citeauthoryear{White}{White}{2005}]{White_2005}
White M.,  2005, Astroparticle Physics, 24, 334–344

\bibitem[\protect\citeauthoryear{Wright, Eisenhardt, Mainzer, Ressler, Cutri,
  Jarrett, Kirkpatrick, Padgett, McMillan, Skrutskie et~al.,}{Wright
  et~al.}{2010}]{wright2010wide}
Wright E.~L.,  Eisenhardt P.~R.,  Mainzer A.~K.,  Ressler M.~E.,  Cutri R.~M.,
  Jarrett T.,  Kirkpatrick J.~D.,  Padgett D.,  McMillan R.~S.,  Skrutskie M.,
    et~al., 2010, The Astronomical Journal, 140, 1868

\bibitem[\protect\citeauthoryear{{Wyithe} \& {Loeb}}{{Wyithe} \&
  {Loeb}}{2009}]{poreion0}
{Wyithe} J.~S.~B.,  {Loeb} A.,  2009, MNRAS, 397, 1926

\bibitem[\protect\citeauthoryear{{Wyithe} \& {Loeb}}{{Wyithe} \&
  {Loeb}}{2007}]{poreion3}
{Wyithe} S.,  {Loeb} A.,  2007, ArXiv e-prints

\bibitem[\protect\citeauthoryear{{Wyithe}, {Loeb} \& {Geil}}{{Wyithe}
  et~al.}{2007}]{param1}
{Wyithe} S.,  {Loeb} A.,    {Geil} P.,  2007, ArXiv e-prints

\bibitem[\protect\citeauthoryear{Yang, Shahalam, Pal, Pan \& Wang}{Yang
  et~al.}{2019}]{yang2019constraints}
Yang W.,  Shahalam M.,  Pal B.,  Pan S.,    Wang A.,  2019, Physical Review D,
  100, 023522

\bibitem[\protect\citeauthoryear{Zafar, Péroux, Popping, Milliard, Deharveng
  \& Frank}{Zafar et~al.}{2013}]{Zafar_2013}
Zafar T.,  Péroux C.,  Popping A.,  Milliard B.,  Deharveng J.-M.,    Frank
  S.,  2013, Astronomy and Astrophysics, 556, A141

\bibitem[\protect\citeauthoryear{Zlatev, Wang \& Steinhardt}{Zlatev
  et~al.}{1999}]{PhysRevLett.82.896}
Zlatev I.,  Wang L.,    Steinhardt P.~J.,  1999, Phys. Rev. Lett., 82, 896

\end{thebibliography}
\end{document}